\def\etal{et~al.}

\documentclass[article]{has40}
\usepackage{graphicx}
\usepackage{amssymb}
\tabletypesize{\normalsize}  

\def\plotone#1{\centering \leavevmode
\includegraphics[width=.55\columnwidth]{#1}}

\def\plotone#1{\centering \leavevmode
\includegraphics[width=.55\columnwidth]{#1}}

\shortauthors{Wilhelm, Spalding, De Lee}
\shorttitle{Panning for RRL Nuggets}

\begin{document}
\large    
\pagenumbering{arabic}
\setcounter{page}{113}

\title{Panning for RRL Nuggets in the SDSS-DR9, Single Epoch Spectra}

%
%
\author{{\noindent Ronald Wilhelm{$^{\rm 1}$}, Eckhart Spalding{$^{\rm 1}$} and Nathan De Lee{$^{\rm 2}$}\\
\\
{\it (1) University of Kentucky, Lexington, KY, USA\\
(2) Vanderbilt University, Nashville, TN, USA} 
}
}

%
%
\email{(1) ron.wilhelm@uky.edu, espalding@uky.edu (2) nathan.delee@vanderbilt.edu }


\begin{abstract}
The SDSS has been a gold mine for understanding properties of the Milky Way.  Below, in the watershed, there remain small nuggets to be found
flowing from the deepest recesses of the mine.
The SDSS-DR9 included the release of the flux- and wavelength-calibrated, individual spectra which were subsequently combined to form the composite
spectra found in previous SDSS data releases.  These single-epoch spectra (SES) can be analyzed to find flux and spectral line variability, and to
probe aspects of phase variations for objects such as RRL stars.  For $\sim 45\%$ of the spectra in the RRL color range, SES were taken on separate nights,
sometimes weeks apart.  The remaining dataset have a time baseline of 0.75 to 1.5 hours and consist of 4-7 separate exposures.   
In my talk I will present details of our project to detect variability and to constrain the pulsation phase at the time of observation.  I will
discuss our auto-detection technique that uses division of all SES for a given star to search for variations in the flux, the hydrogen Balmer lines
and the CaII K line.  This procedure is being tested against known variables and non-variable stars within Stripe 82 to determine the identification
effectiveness as a function of signal-to-noise, RRL type, and time baseline length.  I will also discuss the use of empirical standard star templates
to predict pulsation phase from the combination of hydrogen line strengths and radial velocity variations.  Our ultimate goal is to combine the SES
analysis with our new metallicity calibration in order to increase the number of RRL stars that have reliable metallicity determinations and which
can then be used to probe the structure of the Milky Way halo.
\end{abstract}

\section{Introduction}
The Sloan Digital Sky Survey (SDSS) (York \etal, 2000) and Sloan Extension for Galactic Understanding and Exploration (SEGUE) (Yanny \etal, 2009) have spectral observations of $\sim 660,000$
stars as of Data Release 9 (DR9) (Ahn \etal, 2012).  These spectral observations include tens of thousands of stars that fall in the color range of the RRL instability 
strip, including several thousand RRL variables. These spectra are very useful in determining the RRL radial velocity and can be used to determine the
overall metallicity.  

A typical spectrum from the SDSS database is a composite spectrum produced from multiple single observations.  These single observations sometimes fall on
the same night but often include observations that occur days and even weeks apart.  Combining these observations allows for easy removal of cosmic ray
events and helps to increase the signal-to-noise ratio for the star.  This is a very useful technique if the star is non-variable.  For variable
stars, and in particular the RRL stars, combining single spectra can distort the spectral line profiles, which change as a function of pulsation
phase.  The SDSS-DR9 includes the flux- and wavelength-calibrated, single-epoch spectra (SES) that were used to create the final composite spectrum.  With
this data in hand we now have the ability to measure line strengths that are not distorted from the combining effect and to use the changes in line
strength from multiple SES to identify variables and estimate pulsation phase at the time of observation.

As RRLs pulsate and their surface temperature changes, there is significant change to the strength of absorption lines.  In particular, the
hydrogen Balmer lines and the CaII K line undergo substantial changes as a function of pulsation phase.  In general, the strength of the Balmer
lines are anti-correlated to the strength of the CaII K line, which shrinks as the temperature increases.  This anti-correlation can be exploited in
the SES spectra to identify RRL variables.  Furthermore, the amount of change between SES observations can be used to estimate the phase information 
for the variable.  The phase is of crucial importance for metallicity determination, because at certain pulsation phases---particularly during
light-rise time---shocks in the photosphere cause substantial distortions in the Balmer lines, making it difficult to determine precise metallicity.   
 
In this talk we present results for our identification and phase estimation technique by comparing to known SDSS variables in Stripe 82 (Sesar \etal, 2010). Here 
we constrain the test to RRab variables and to SES that are taken in a single night.  Future testing will include RRc variables and stars with SES
spectra taken on multiple nights.

\pagebreak
\section{Stripe82 variables and the Empirical Templates}
We have chosen eleven RRab variables from Stripe 82, with phase information reported in the Sesar paper.  We also chose five stars
that are non-variable to test our ability to detect variability.  All test stars have SES taken on a single night with observations taken
consecutively.  The number of SES range from a minimum of three spectra up to seven.  Table 1 lists these stars, their period, the time base line of
observation (T-Base), the percent of the phase covered by the SES, and the magnitude of the star.

The ability to detect variation will be a function of T-Base line, the period of the variable, and the signal-to-noise of the SES spectra.  A further
constraint is the pulsation phase during the observation because changes in line strengths occur more rapidly at certain pulsation phases.  For the
non-variable stars, we chose SES that covered similar ranges in T-Base
and signal-to-noise to test for null detections. 
\begin{flushleft}
\begin{deluxetable*}{rccccc}
\tabletypesize{\normalsize}
\tablecaption{Stripe82 Test Sample}
\tablewidth{0pt}
\tablehead{ \\ \colhead{Spectrum}   & \colhead{Class} &
       \colhead{Period (Days)} & \colhead{T-Base (hours)} & \colhead{Percent of Phase} &
  \colhead{g-mag} \\
}

\startdata
 0989-52468-0032 &  ab  & 0.580767811 &  0.7139 &  0.0512  & 18.388 \\
 0702-52178-0100 &  ab  & 0.649984725 &  0.7522 &  0.0482  & 17.845 \\
 0402-51793-0117 &  ab  & 0.797848004 &  0.6364 &  0.0332  & 15.763 \\
 1918-53240-0121 &  ab  & 0.512230079 &  0.6528 &  0.0531  & 14.782 \\
 1507-53763-0132 &  ab  & 0.722990487 &  0.6308 &  0.0364  & 17.946 \\
 1502-53741-0170 &  ab  & ...         &  1.7386 &  0.1200  & 16.383 \\
 0380-51792-0237 &  ab  & 0.569410914 &  0.8797 &  0.0644  & 18.204 \\
 2636-54082-0248 &  ab  & 0.542616712 &  2.2861 &  0.1755  & 17.998 \\
 0371-52078-0334 &  ab  & 0.622160962 &  0.6355 &  0.0425  & 16.913 \\
 0685-52203-0526 &  ab  & 0.793004216 &  1.0889 &  0.0572  & 18.850 \\
 0701-52179-0570 &  ab  & 0.557696402 &  0.6147 &  0.0459  & 17.693 \\
 1903-53357-0185 &  nv  & ...         &  1.3319 &  ...     & ...   \\
 1918-53240-0241 &  nv  & ...         &  0.6528 &  ...     & ...   \\
 1084-52591-0422 &  nv  & ...         &  0.7633 &  ...     & ...   \\
 1138-53228-0444 &  nv  & ...         &  1.1444 &  ...     & ...   \\
 1133-52993-0463 &  nv  & ...         &  1.0608 &  ...     & ...   
\enddata
\end{deluxetable*}
\end{flushleft}

\vskip -0.4in
To test for variability, we use empirical templates obtained from observations of metallicity standard RRab stars.  These spectra were taken at 
McDonald Observatory on the 2.1meter telescope using the es2, low-resolution spectrograph.  These stars are being used in our metallicity calibration,
which has been reported at this meeting by E. Spalding.  The spectra were taken over large portions of the pulsation phase.  Currently none of the stars 
have complete phase coverage, so for this study we have used a combination of templates from the variable stars X Ari, TT Lyn, TW Lyn and TU Uma.
The combination of these stars gives full phase coverage with a phase resolution of $10\%$.

In our technique we make use of the quotient spectrum, the division of one spectrum by another, to characterize the changes in spectral line strength
in a given phase range.  Because our current template sample is constructed from stars with various metallicities, we only divide spectra for a given
star to construct the phase range.  This minimizes effects of metallicity, particularly on the CaII K line, for the quotient templates. 

Figures 1 and 2 show two quotient templates at different phases.  The quotient spectra would have a value of unity, across all wavelengths, if there were no
spectral changes.  In figure 1 the hydrogen Balmer lines can be seen in ``absorption'' (decreasing line strength) while the CaII K line shows an ``emission'' peak
(increasing line strength).  The blend of CaII H/H{$\epsilon$} 
 displays broad wings and an ``emission'' peak at the line center, clearly showing the anti-correlated nature of the CaII lines and the Balmer lines.  In figure 2, during 
the light-rise phase, the effect is reversed with the Balmer lines growing while the CaII lines are shrinking.  Comparing these two figures, it becomes evident how the phase can be estimated, based on the sign and strength 
of the spectral features from the quotient templates.

To test our technique we make a quotient spectrum from the division of the first and final SES spectra for a given star.  A particularly good result can 
be seen in figure 3.  Here the black line is the quotient spectrum for a Stripe82 variable, while the red line is the best matched quotient template.  It is clear 
in this figure that the star is a variable and at a phase approaching 1.0.  

\begin{figure*}
\centering
\plotone{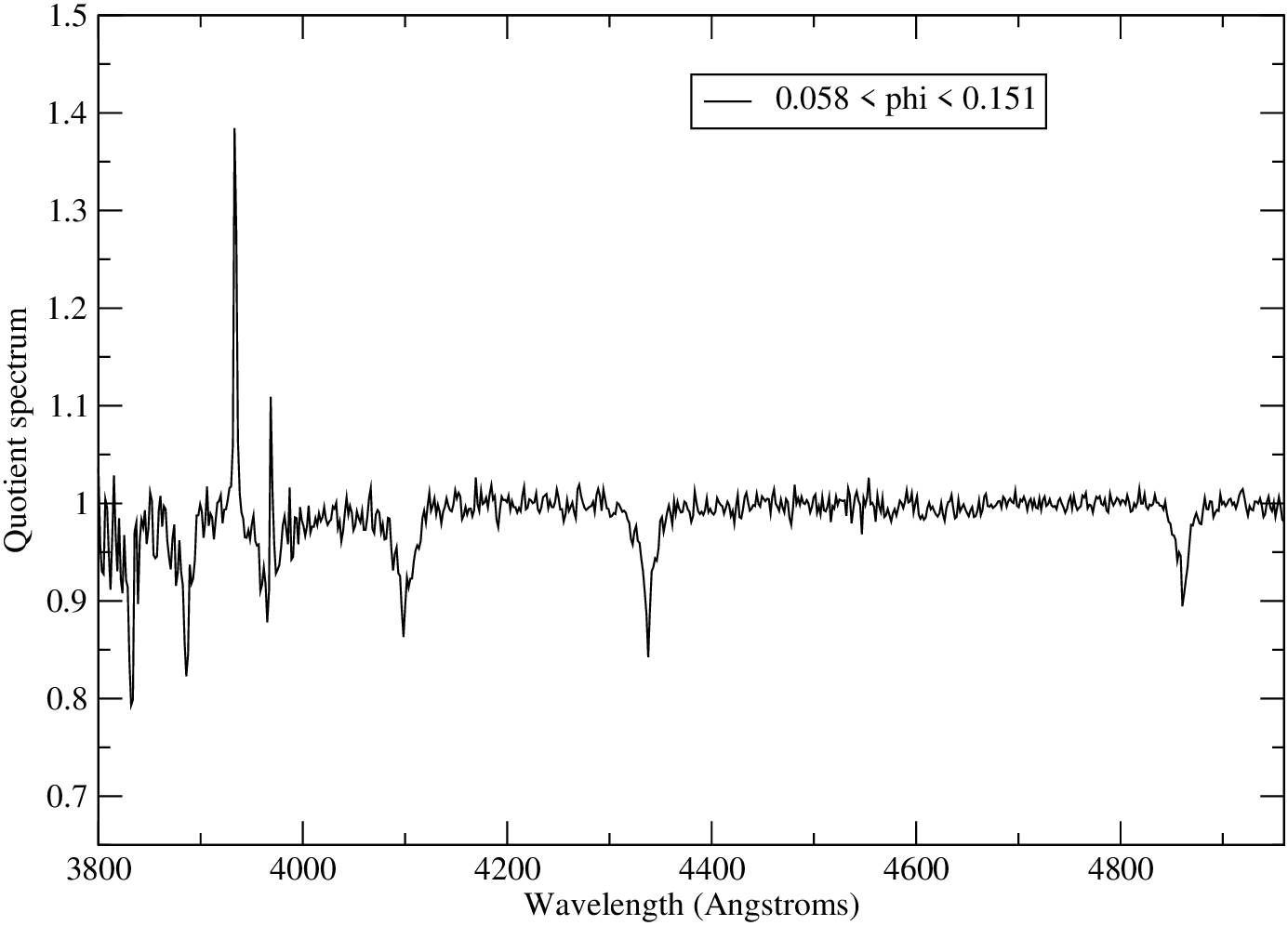}
\vskip 12pt
\caption{Template quotient spectrum showing the division of two spectra taken at a phase of 0.0 and 0.1.  The continuum is centered on unity while the hydrogen lines
have values less than unity, due to a decrease in line strength.  The CaII K line is above unity, indicating that the CaII K line is growing in strength.  The "emission" 
peak near the core of the H{$\epsilon$} line at 3966 Angstroms is due to the increasing strength of the CaII H line.}
\label{1}
\end{figure*}

\begin{figure}
\centering
\plotone{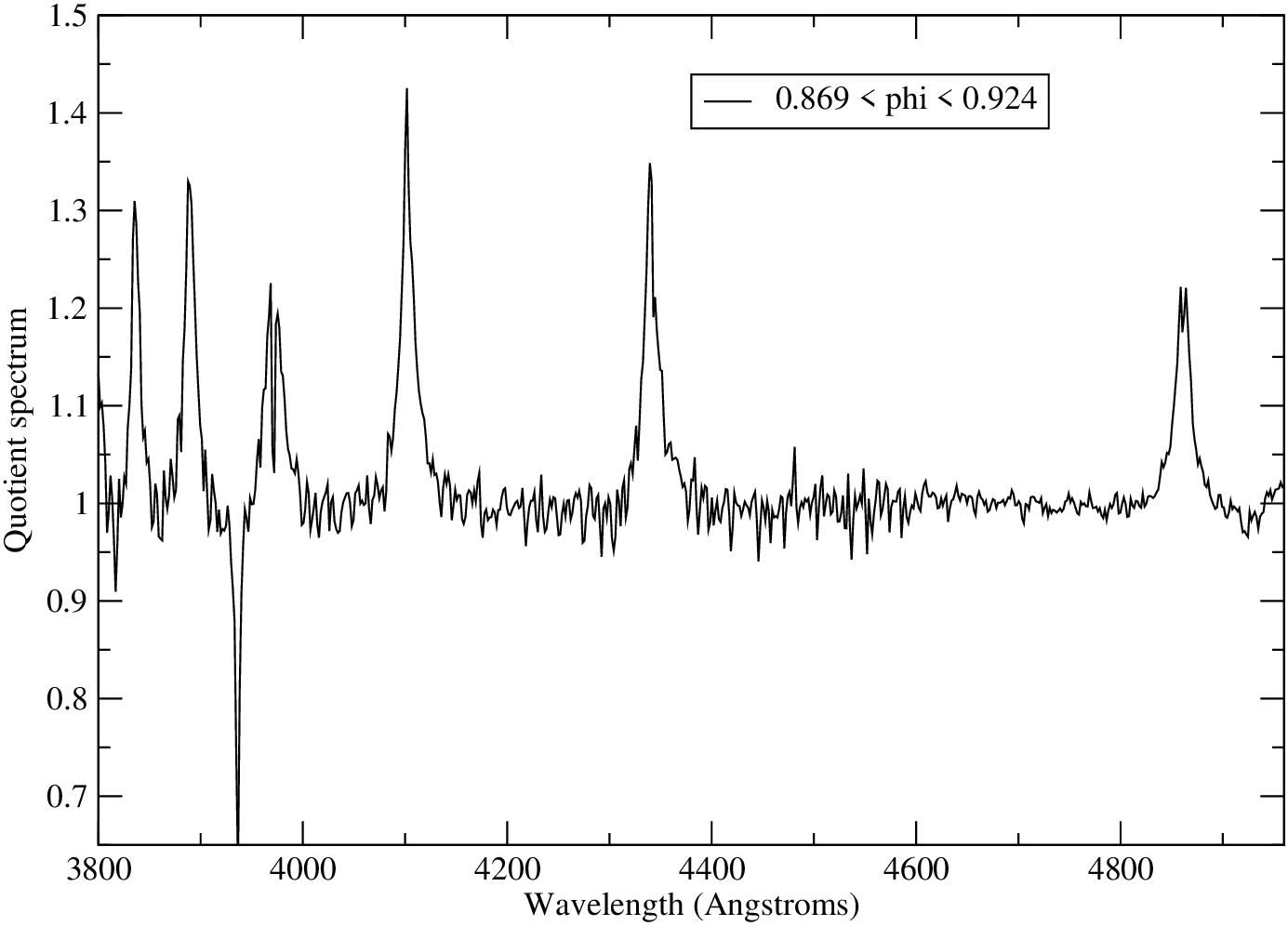}
\vskip 12pt
\caption{Same as figure 1 but for a phase between 0.8 and 0.9.  The hydrogen lines are above unity indicating line growth, while the Ca II K line is shrinking.  Again the signature
of the CaII H line is clear in the H{$\epsilon$} peak. }
\vskip 12pt
\label{2}
\end{figure}

\begin{figure}
\centering
\plotone{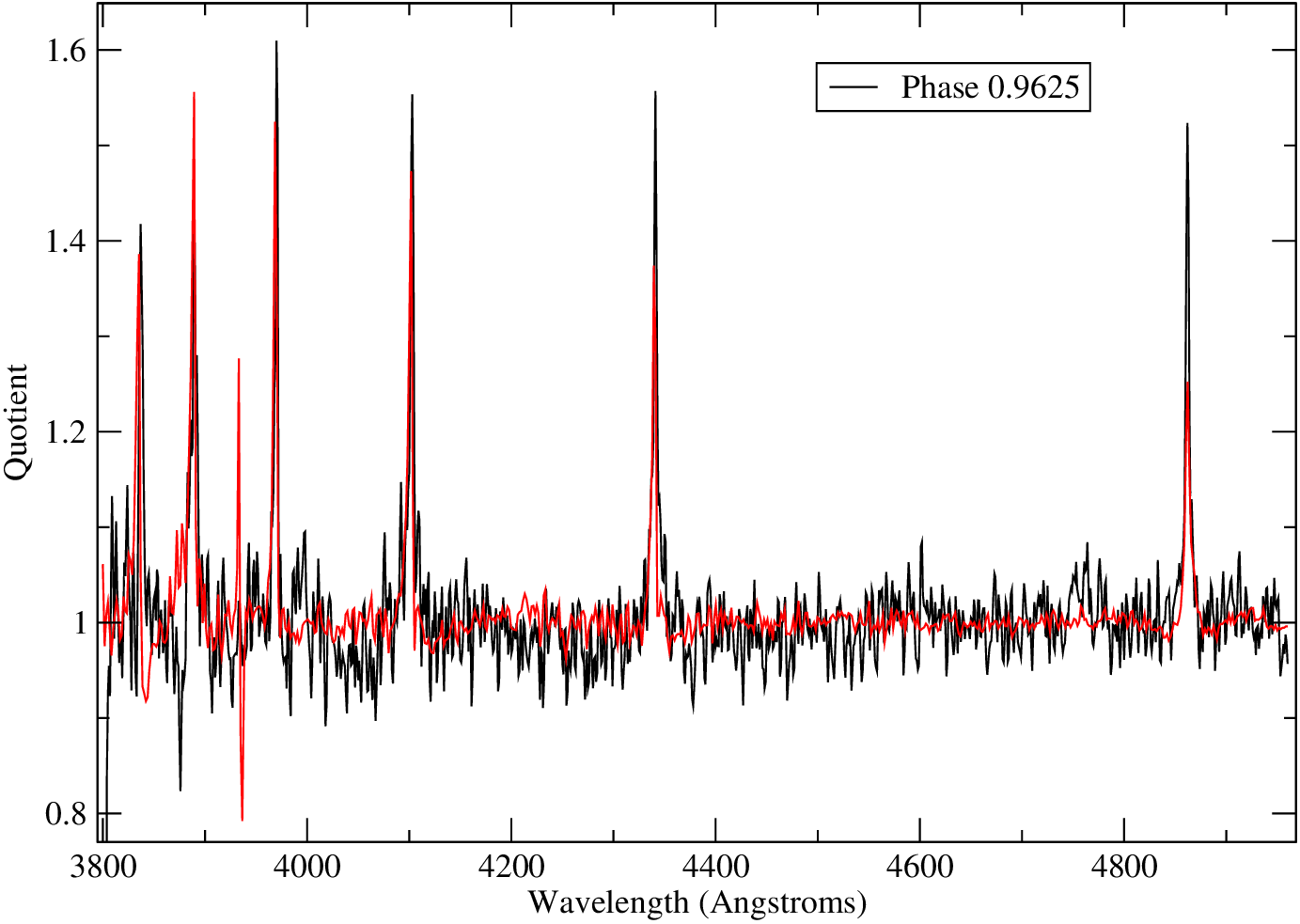}
\vskip 12pt
\caption{The quotient spectrum for an RRab star in Stripe82.  The signal is unusually strong for this particular star.  The red line is the best template match to the data.  It is clear that this 
star is not only a variable, but is consistent with the phase of the template. }
\label{3}
\end{figure}

\section{The Technique}

To quantify the signal for the variable candidates we perform a FFT (IRAF task fxcor) to correlate the template and observed quotient spectra.  We use a 
Welch Fourier filter with a wavenumber range of $10 < k < 500$, where the low frequency cutoff removes large scale variations produced by the flux 
normalization and a high frequency limit to remove high frequency noise.  We further constrained our correlation peak to the velocity range of 
$-800 < v < 800$ km/s, to eliminate correlation peaks that would have unrealistic velocities for stars in the Galaxy.  The Tonry-Davis ratio, which is the ratio of correlation height to anti-symmetric noise (Tonry \& Davis, 1979), is used to 
determine the significance of the correlation peak and to select the best quotient template to match the observed data.

Figure 4 shows the TDR value as a function of various phase templates.  The filled black circles are the Stripe82 variable and the black-hatched squares are
the auto-correlation for the $\phi =  0.960$ template relative to all template spectra.  The red circles and red squares are the anti-correlation responses 
for the Stripe82 star and the template, respectively.  The anti-correlation is found to be inverting the templates, and thus the greater the negative peak, the stronger
the anti-correlation.  It is clear that the Stripe82 star has virtually the same template response as the template with $\phi = 0.960$.  This increases the 
likelihood that the TDR peak is real and not a random variation.

As a final test, we use the velocities from the star's spectra, and test if the velocity from the peak correlation of the quotient spectra, yield a consistent result.
For this strong signal star it is clear that the TDR peak occurs at the velocities of the individual spectra (figure 5).  In cases where the TDR peak is weak and only
due to noise variations, it is 
less likely that the velocity of the correlation peak will match the velocity of the single spectra.  This constraint adds more certainty 
to the significance of the TDR peak.  Both the auto-correlation and the radial velocity constraints must be met in order
for the variability result to be considered positive.

\begin{figure*}
\centering
\plotone{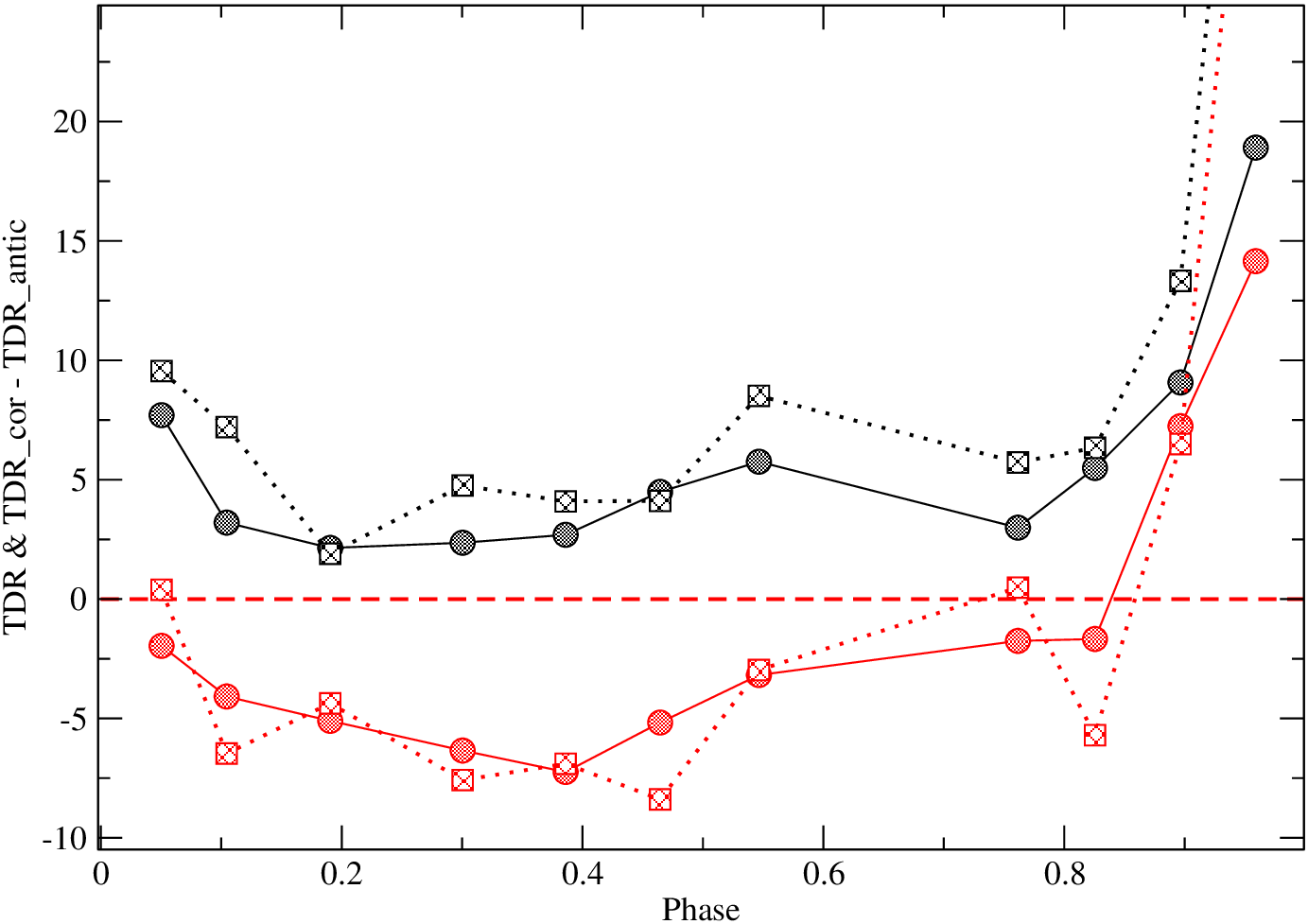}
\vskip 13pt
\caption{Comparison of the TDR response with phase for the Stripe82 variable, filled circles and the auto-correlation response, hatched squares.  The black symbols are for the correlation peaks and
the red are the anti-correlation peaks.  It is clear that for this variable, the responses are in keeping with the auto-correlation result. }
\label{4}
\end{figure*}

\begin{figure*}
\centering
\plotone{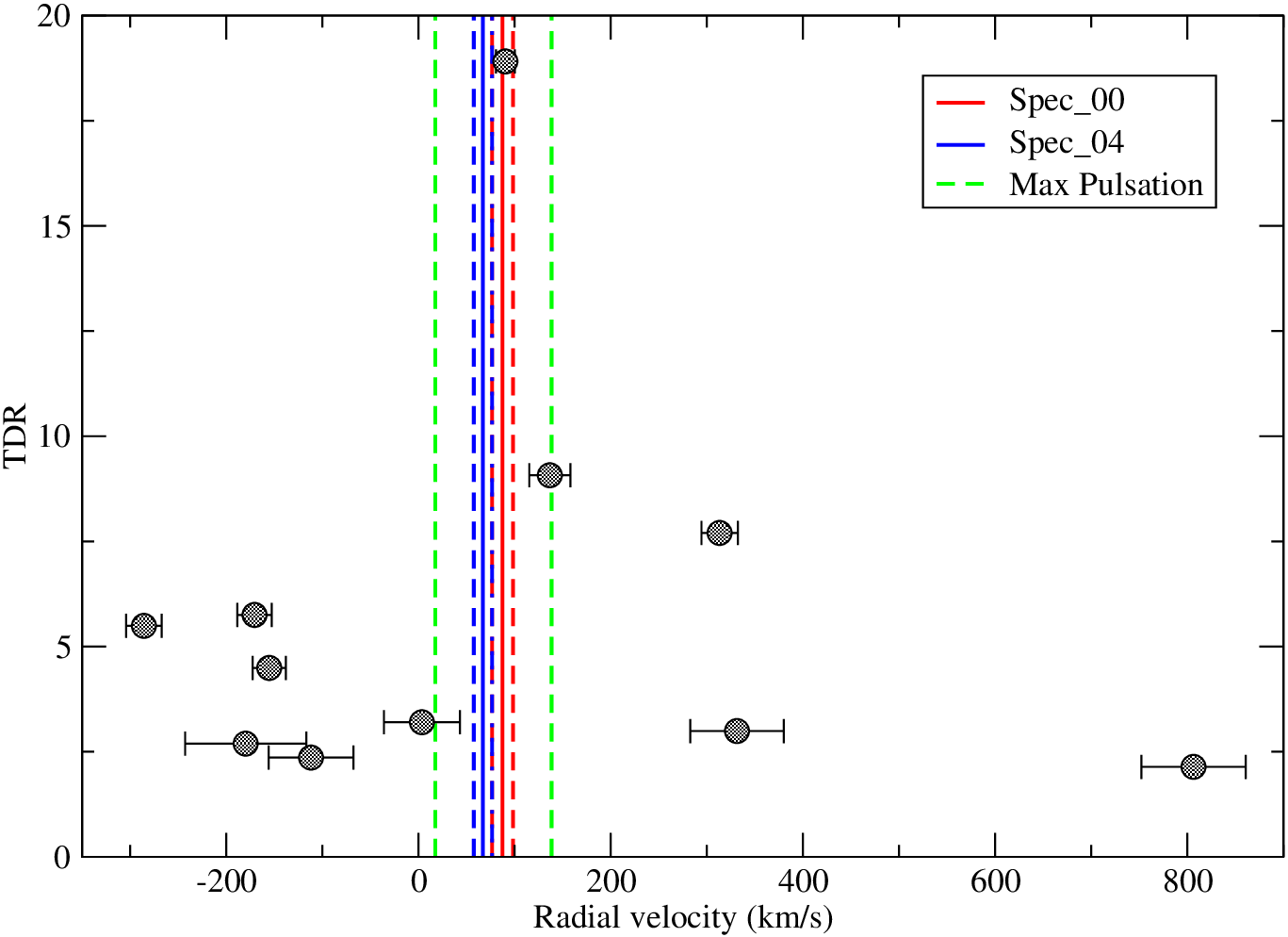}
\vskip 12pt
\caption{The TDR values verse the velocity for the correlation peak.  The solid and dashed lines represent the velocity values and error from the single spectra that make up the quotient spectrum.  It is evident in this figure that the peak correlation and single spectra velocities are consistent. }
\label{5}
\end{figure*}

\pagebreak
\section{The Results}
Of the eleven RRab stars that we tested, only four were found to have a significant TDR peak.  The other seven variables failed either the auto-response criteria, the 
radial velocity criteria, or both.  Of the six non-variable stars that we tested, all six failed the variability test.  Figure 6 is a plot of the TDR peak height as a 
function of the time baseline.  The black circles are the positive detections and the red circles are the null results.  The green squares represent the non-variable 
stars.  In general the null RRab detections are for time baselines that are less than 45 minutes, indicating that the time baseline length is an important criteria for 
variability detection.  The low peak height of the non-variable stars shows that the height of the TDR peak is an indicator of variability, as expected.

Figure 7 shows the results for the predicted phase versus the phase computed from the ephemeris data of Sesar.  The four positive detections are close to the one-to-one
correspondence line, although there appears to be a systematic shift for the two stars with the smallest phase.  The null detections appear more randomly distributed 
in the plot.  This suggests that for positive detections it will be possible to estimate the phase to within $10\%$ or so.  It should be noted, however, that the positive
detections were found to be during the phases close to maximum light, when the spectral changes occur the most rapidly.  This suggests that detections are most likely
to be found during phases of rapid spectral line change.  This is again expected, given the short time baseline for most of these stars.

\begin{figure*}
\centering
\plotone{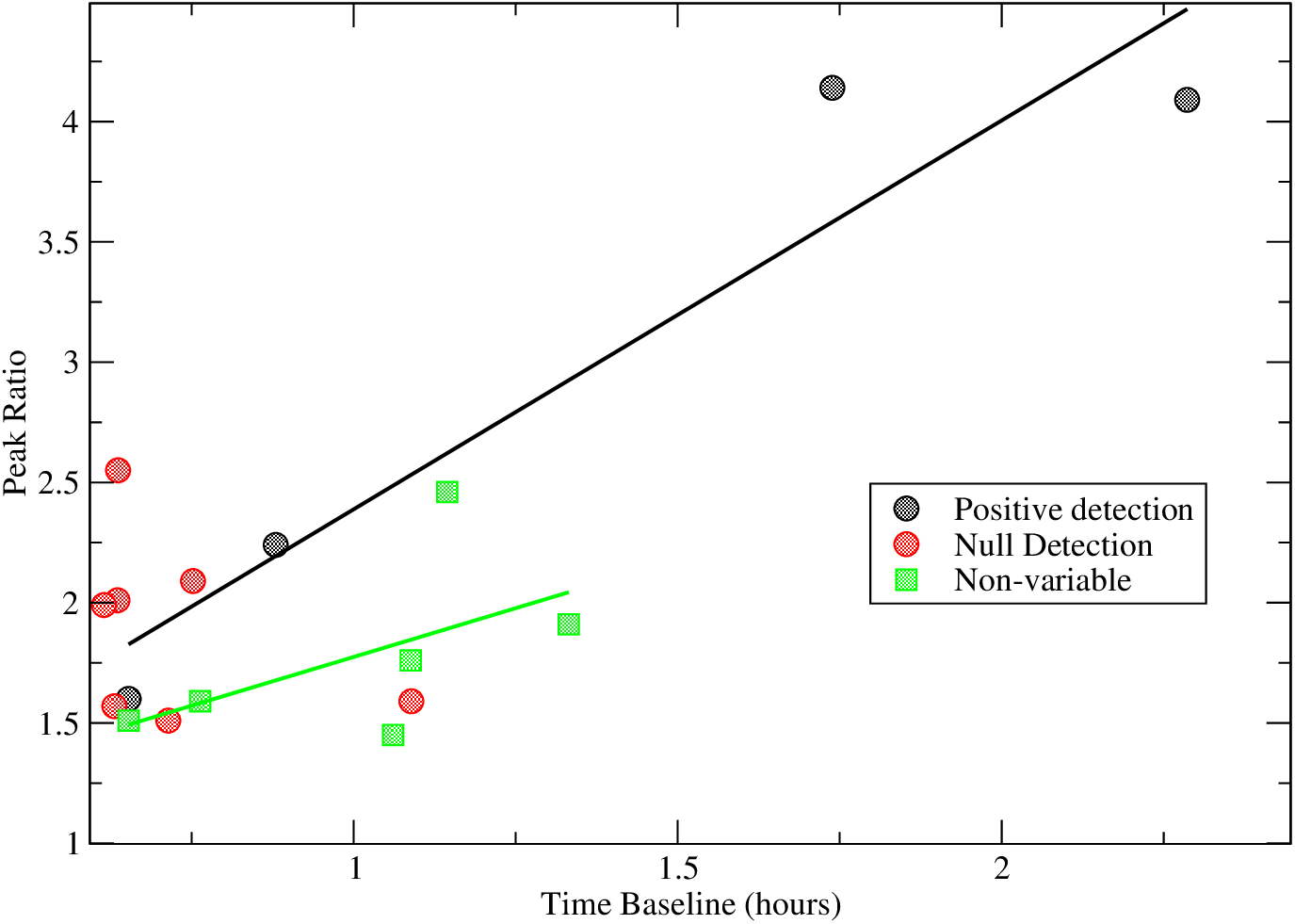}
\vskip 12pt
\caption{Comparison of Peak TDR value versus the time baseline of the observations. Positive detections are solid black circles, null detections are red circles and the 
green squares are null detections for the non-variable sample.  The plot shows that longer baseline observations and larger TDR peaks have a better possibility for detection. }
\vskip 12pt
\label{6}
\end{figure*}

\begin{figure*}
\centering
\plotone{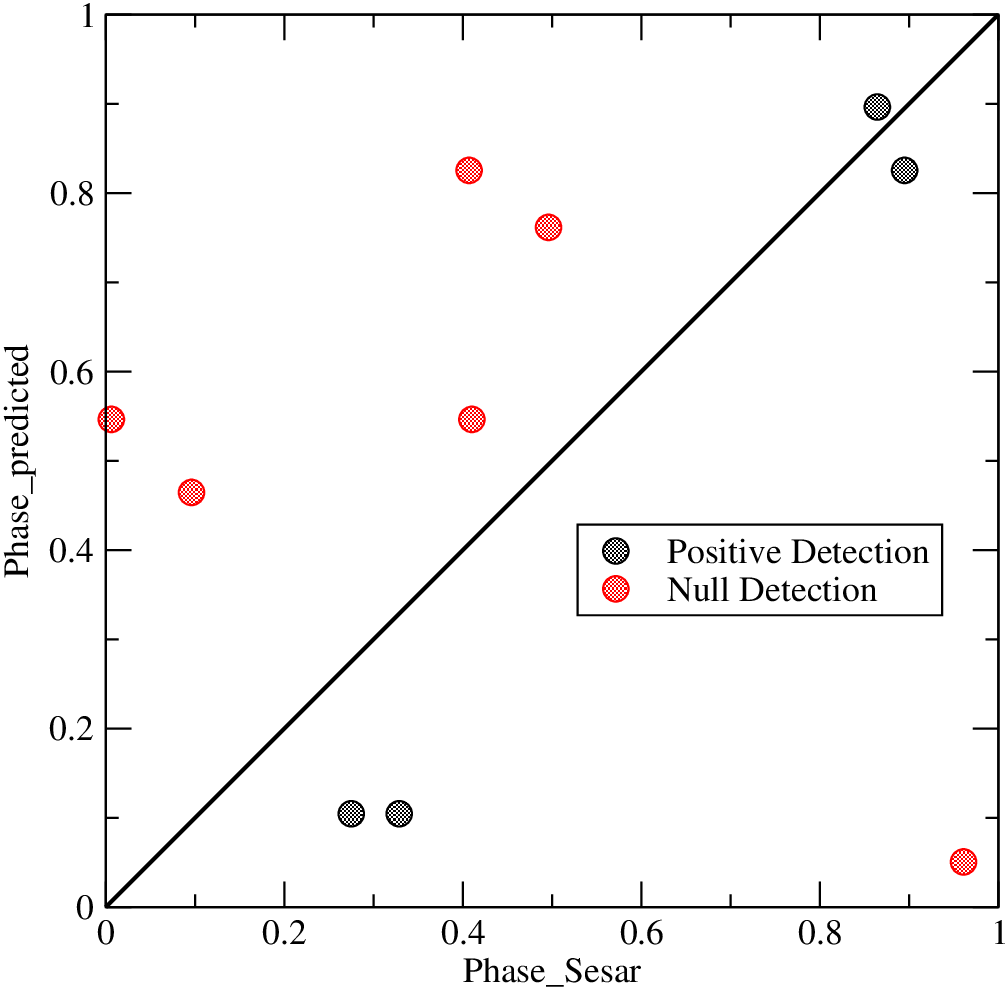}
\vskip 20pt
\caption{Comparison of phase estimations versus the phase from Sesar ephemerides.  The positive detections are close to the one-to-one correspondance line while the null detections
are more randomly distributed in the plot.  Note that only observations near maximum light phase show positive detections. }
\label{7}
\end{figure*}

\pagebreak
\section{Conclusion}

We find that using our empirical templates it is possible to detect variability and estimate phase for RRab stars.  When the SES are all constrained to a single night, the 
success rate is mainly determined by the time baseline and, in particular, the percentage of the pulsation period that is represented by the data.  For all the positive 
detections, a time baseline that is greater than $5\%$ of the pulsation period is needed to return a positive result.  It is also most likely that positive results will 
occur for phases near the maximum light, when the spectral changes are occurring rapidly.  We see no strong relation between positive detections and the signal-to-noise ratio of the spectra. 
This suggests that the technique will work for very low signal-to-noise spectra.

We have not tested the RRc variables at this point, but the shorter periods will make even the shortest time baselines a large percentage ($ > 10\%$) of the pulsation 
phase.  This should increase in the number of positive detections for single-night, SES, although the spectral changes are not as extreme as the RRab stars.  We have also 
not tested the $46\%$ of the stars that have SES taken on multiple nights.  For these we expect a much higher frequency of positive detections, but the phase determination will
be more difficult due to an unknown number of pulsation phases between the observations.

This July we are continuing the observation of standard RRL variables.  Our goal is to observe an RRab and RRc through the entire pulsation period and at high time resolution (
$ < 5\%$ of pulsation period)  This should increase the precision of the phase estimation.

\end{document}